%% file: main.tex
\documentclass[conference]{IEEEtran}
\IEEEoverridecommandlockouts
\usepackage[numbers]{natbib}
\usepackage[switch]{lineno} 
\usepackage{adjustbox}
\usepackage{tabularx}
\usepackage{multirow}
\usepackage{caption}
\usepackage{subcaption}
\usepackage{svg}
\usepackage{makecell}
\usepackage{graphicx}
\usepackage{booktabs}
\usepackage{caption}
\usepackage{float}
\usepackage{hyperref}
\usepackage{color}
\usepackage{tikz}
\usepackage{amssymb}
\usepackage{pifont}
\usepackage{xcolor}
\usepackage{pdflscape}
\usepackage{afterpage}
\usepackage{capt-of}
\usepackage{comment}
\usepackage{tikz,colortbl}
\usetikzlibrary{calc}
\usepackage{zref-savepos}
\usepackage{bbding}
\usepackage{hyperref}
\usepackage{amsmath,amssymb,amsfonts}
\usepackage{algorithmic}

\newcommand{\tikzcircle}[2][red,fill=red]{\tikz[baseline=-0.5ex]\draw[#1,radius=#2] (0,0) circle ;}%

\usepackage{listings}
\usepackage{xcolor}

\definecolor{codegreen}{rgb}{0,0.6,0}
\definecolor{codegray}{rgb}{0.5,0.5,0.5}
\definecolor{codepurple}{rgb}{0.58,0,0.82}
\definecolor{backcolour}{rgb}{0.95,0.95,0.92}

\lstdefinelanguage{Solidity}{
  keywords={contract, struct, modifier, break, case, catch, continue, debugger, default, delete, do, else, finally, for, function, if, in, instanceof, new, return, switch, throw, try, typeof, var, void, while, with},
  keywordstyle=\color{red}\bfseries,
  ndkeywords={class, export, boolean, throw, string, address, import, this},
  ndkeywordstyle=\color{blue},
  identifierstyle=\color{black},
  sensitive=false,
  comment=[l]{//},
  morecomment=[s]{/*}{*/},
  commentstyle=\color{codegreen}\ttfamily,
  stringstyle=\color{codepurple}\ttfamily,
  morestring=[b]',
  morestring=[b]"
}

\usepackage[linesnumbered,ruled,vlined]{algorithm2e}
\SetKwInput{KwInput}{Input}                
\SetKwInput{KwOutput}{Output}              
\SetKwInput{KwFunction}{Function}              

\def\BibTeX{{\rm B\kern-.05em{\sc i\kern-.025em b}\kern-.08em
    T\kern-.1667em\lower.7ex\hbox{E}\kern-.125emX}}

\begin{document}

    \makeatletter
    \newcommand{\linebreakand}{%
      \end{@IEEEauthorhalign}
      \hfill\mbox{}\par
      \mbox{}\hfill\begin{@IEEEauthorhalign}
    }
    \makeatother

\let\WriteBookmarks\relax
\def\floatpagepagefraction{1}
\def\textpagefraction{.001}



\title{D-VRE: From a Jupyter-enabled Private Research Environment to Decentralized Collaborative Research Ecosystem

}

\author{
    \IEEEauthorblockN{Yuandou Wang\IEEEauthorrefmark{1}, Sheejan Tripathi\IEEEauthorrefmark{1}\IEEEauthorrefmark{2}, Siamak Farshidi\IEEEauthorrefmark{1}, Zhiming Zhao\IEEEauthorrefmark{1}\IEEEauthorrefmark{3}}
    \IEEEauthorblockA{\IEEEauthorrefmark{1}Multiscale Networked System, University of Amsterdam, The Netherlands
    }
    \IEEEauthorblockA{\IEEEauthorrefmark{2}Department of Computer Science, Vrije Universiteit Amsterdam, The Netherlands
    }
    \IEEEauthorblockA{\IEEEauthorrefmark{3}LifeWatch ERIC Virtual Lab and Innovation Center, Amsterdam, The Netherlands    
    }
    
}    

\maketitle

\begin{abstract}
Today, scientific research is increasingly data-centric and compute-intensive, relying on data and models across distributed sources. However, it still faces challenges in the traditional cooperation mode, due to the high storage and computing cost, geo-location barriers, and local confidentiality regulations. The Jupyter environment has recently emerged and evolved as a vital virtual research environment for scientific computing, which researchers can use to scale computational analyses up to larger datasets and high-performance computing resources. Nevertheless, existing approaches lack robust support of a decentralized cooperation mode to unlock the full potential of decentralized collaborative scientific research, e.g., seamlessly secure data sharing. In this work, we change the basic structure and legacy norms of current research environments via the seamless integration of Jupyter with Ethereum blockchain capabilities. As such, it creates a \underline{D}ecentralized \underline{V}irtual \underline{R}esearch \underline{E}nvironment (D-VRE) from private computational notebooks to decentralized collaborative research ecosystem. We propose a novel architecture for the D-VRE and prototype some essential D-VRE elements for enabling secure data sharing with decentralized identity, user-centric agreement-making, membership, and research asset management. To validate our method, we conducted an experimental study to test all functionalities of D-VRE smart contracts and their gas consumption. In addition, we deployed the D-VRE prototype on a test net of the Ethereum blockchain for demonstration. The feedback from the studies showcases the current prototype's usability, ease of use, and potential and suggests further improvements. 
\end{abstract}


\maketitle

\section{Introduction}

In the contemporary era, data is at the center of most scientific research, particularly in fields such as medical research~\citep{sidey2019machine}, climate science~\citep{monteleoni2013climate}, and biodiversity~\citep{mathur2023machine}, which usually require data and analysis models across distributed sources. To tackle complex scientific problems, teams of researchers used to travel to different institutes, reserving time slots to discuss collaborations, collect data, and then analyze the data afterward~\citep{henderson2020accelerating}. 


The Jupyter\footnote{\url{https://jupyter.org/}}, well-known as data scientists' computational notebook of choice, has played a vital facilitating role in such processes for scientific computing~\citep{perkel2018jupyter}. 
On the one hand, Jupyter is already well-established in the scientific toolbox for small-scale, single-node exploratory analysis with the Jupyter Notebook interface. JupyterLab\footnote{\url{https://github.com/jupyterlab/jupyterlab}} can support custom applications and extensions that live alongside the core Notebook interface. JupyterHub\footnote{\url{https://github.com/jupyterhub/jupyterhub}} can serve a variety of research environments that can be configured with authentication to provide access to a subset of users to share data and scientific workflows remotely. On the other hand, Jupyter shares some features with Virtual Research Environments (VREs) and Science Gateways (SGs), for example, they are Web-based platforms providing scientists with computational services. Over the past couple of years, many studies have explored how it would be beneficial to scale computational analyses up on larger datasets and compute resources using Jupyter~\citep{gazzarrini2024virtual, zhao2022notebook,Assante2023, wang2013cybergis, Yin2017, zonca2018deploying,lawrence2015science}. It is known that the current success of the traditional cooperation mode should be attributed to the centralized research data lifecycle governance pattern~\citep{huang2019software} that usually covers identity management, sharing policies, research assets transfer, workflow orchestration, and infrastructure management. There are a few works in which the sharing policies are in whole or part based upon terms defined by the end-users. For example, D4Science~\citep{assante2019enacting} allows users to define authorization policies that satisfy the authorization requirements using the policy language. WorkflowHub~\citep{goble2021implementing}, based on the FAIRDOM-SEEK framework\footnote{\url{https://docs.seek4science.org/}, \url{https://fair-dom.org/}}, allows users to choose different levels of sharing permissions pre-defined in the framework. 
The fact is that the traditional centralized cooperation mode is not a preferable choice today, with the advent of the explosive burst of big data. The fully centralized pattern has inherent disadvantages, including increased data traffic and concerns about data ownership, privacy, security, and confidentiality~\citep{warnat2021swarm}. Undoubtedly, the increasing data volumes significantly require large storage and computing costs in the centralized management paradigm~\citep{blythman2022libraries}. Noticeably, data assets are naturally decentralized and stored on geo-graphically distributed infrastructures or facilities. Data policies and compliance require researchers or data scientists to include robust and detailed plans for how the research data will be managed and shared during the entire lifecycle of their research. Therefore, it still faces challenges in research asset management due to geo-location barriers and local confidential regulations. 

We hypothesize that the decentralized cooperation mode would overcome the current weaknesses of centralized ones, and accommodate inherently decentralized data and sharing policies regarding confidentiality and workflow efficiency in the lifecycle of scientific research. In the past few years, impressive progress has been achieved by decentralized models using blockchain~\citep{zheng2018blockchain, zheng2017overview} that remove the central control elements from the system and decentralize its role for various purposes, e.g., enabling controllable remote computing for sharing based on a zero-trust decentralized identity infrastructure~\citep{rong2022openiac} and confidential distributed machine learning~\citep{warnat2021swarm}. Decentralized Science (DeSci) and Decentralized Artificial Intelligence (DeAI) are hot topics emerging with the advent of the infrastructure of Web 3.0 in recent years~\citep{blythman2022libraries, huang2019software, ding2022desci, cao2022decentralized}. 
DeSci~\citep{ding2022desci} is an attempt to solve the bottleneck problems about oligarchy and data silos, and make science more fair, accessible, responsible, and sensitive. Blythman \textit{et al.}~\citep{blythman2022libraries} explore the potential of decentralized technologies for DeAI to address the limitations caused by current centralized solutions. However, since work is still in its very early stages, not all aspects are managed in a decentralized manner for collaborative research. 

Our prior work~\citep{zhao2022notebook, wang2022scaling} has demonstrated how Jupyter Notebook as a VRE can provide user-centric support in the lifecycle of research activities, e.g., discovering and accessing research assets or composing application workflows, and executing them on Cloud infrastructures. However, it still lacks built-in decentralized cooperation capabilities to speak various decentralized computing platforms, applications, or custom sharing policies. In this article, we target leveraging decentralized technologies and Jupyter to bridge this gap and propose a decentralized VRE (D-VRE) framework to serve as a reference for the development and implementation of such a decentralized cooperation mode in Jupyter-enabled virtual research environments. 
The main contribution of this work can be summarized as follows: 
\begin{itemize}
    \item We characterize existing research environment exemplars by comparing the management of identity, asset, and infrastructure, as well as workflow orchestration. 
    
    \item We propose the D-VRE framework for decentralized collaborative research within a human-in-the-loop scientific research lifecycle context. 
    
    \item We propose a detailed architecture design and description of the D-VRE components and the custom smart contract-based sharing policy as its core functionality.  
    
    \item We provide a Proof-of-Concept (PoC) implementation of the D-VRE architecture, and its evaluation by an experimental study and a case study. 

\end{itemize}

The reminder of this article is structured as follows: Section~\ref{sec:related} review the current state of existing VRE-related exemplars and analyzes research gaps. Section~\ref{sec:D-VRE} presents the conceptual framework, architecture design, and technology choices of the proposed D-VRE. Section~\ref{sec:prototype} details the PoC implementation, experimental and case studies, and its evaluation. In Section~\ref{sec:discussion}, we discuss open challenges and future improvements; and finally, Section~\ref{sec:conclusion} concludes this work.

\section{Related Works}\label{sec:related}

This section characterizes the existing works into two dimensions of unity of opposites: centralized vs. decentralized management of assets and co-located vs. distributed infrastructure resources, as presented in the axis of Figure~\ref{fig:class}. To make the analysis deeper, we focus on the four main aspects: (\tikzcircle[fill=green]{4pt}) identity management, (\tikzcircle[fill=orange]{4pt}) sharing policy, (\tikzcircle[fill=gray]{4pt}) asset transfer, and (\tikzcircle[fill=blue]{4pt}) workflow orchestration. 
\begin{figure*}
    \centering
    \includegraphics[width=\linewidth]{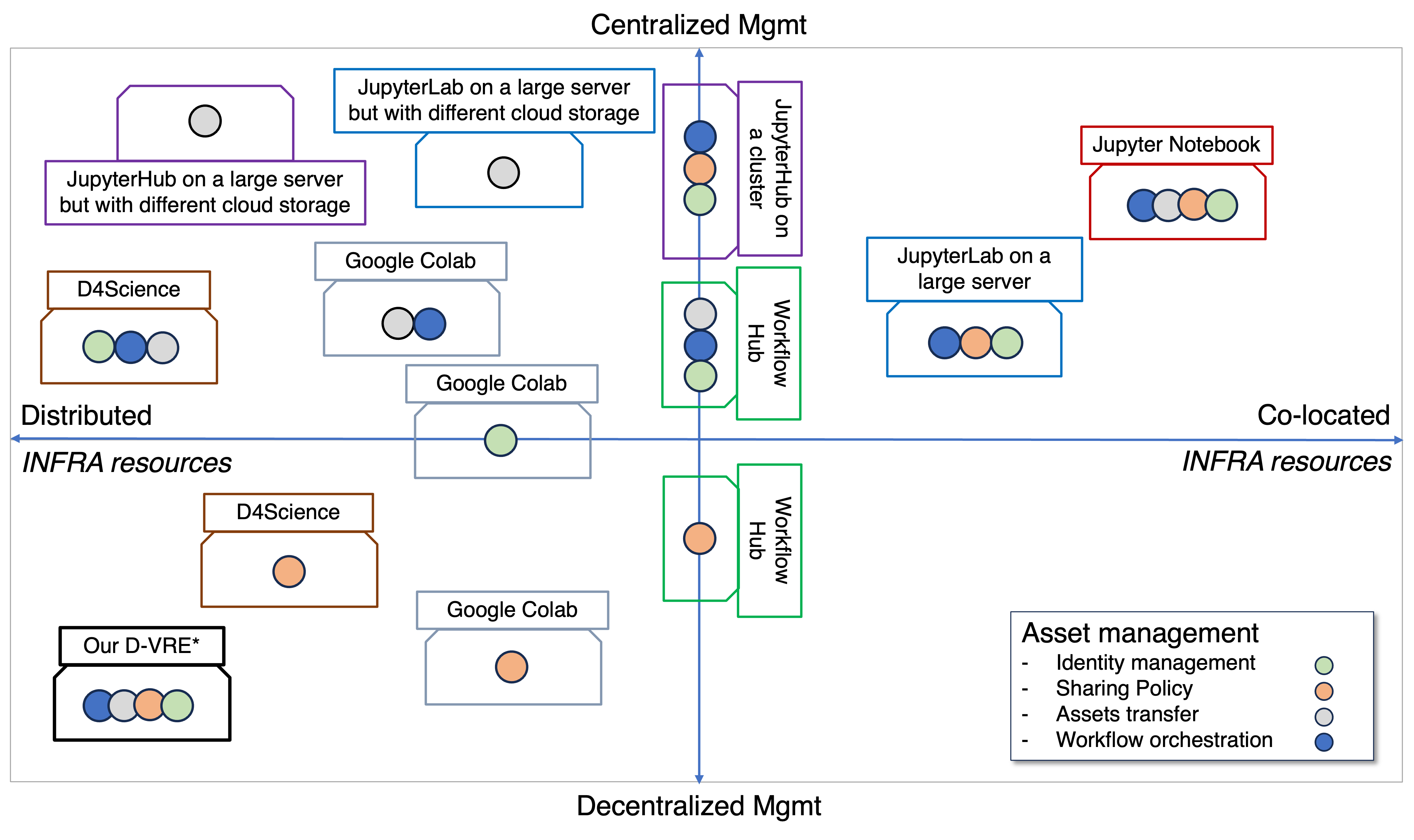}
    \caption{An overview of comparisons of existing research environment exemplars by comparing the ways of identity management (\tikzcircle[fill=green]{4pt}), sharing policy (\tikzcircle[fill=orange]{4pt}), assets transfer (\tikzcircle[fill=gray]{4pt}), and workflow orchestration (\tikzcircle[fill=blue]{4pt}).}
    \label{fig:class}
\end{figure*}

Jupyter Notebook, as a computational notebook, is typically used for small-scale and single-node analysis, such as on personal computers or co-located infrastructure. 
Over the past few years, several studies have implemented interactive large-scale services with Jupyter to scale computational analyses on HPC, cloud, and hybrid HPC-Cloud scientific computing facilities. These approaches utilize Jupyter as a companion application to gateways, lowering the entry barrier to various computational resources on large servers, clusters, or distributed infrastructures.
Examples of these implementations include the CyberGIS-Jupyter framework~\citep{Yin2017}, HPC Notebooks services~\citep{milligan2018jupyter}, Jupyter at NERSC~\citep{henderson2020accelerating}, and deployment of JupyterHub on XSEDE resources~\citep{zonca2018deploying}. Although these works support assets (e.g., data and/ or analyses) transfers across distributed sources, management remains centralized as the computing center typically handles the operations and maintenance of the platform. Besides, the authentication scheme was usually implemented by reverse HTTP proxy with third-party providers for authentication. In some instances, the back-end comprises a federated computing system involving multiple organizations. Nevertheless, the decision-making layer remains centralized, relying on the consensus mechanism among the participating parties.

Apart from the deployment of the Jupyter environment on HPC, VRE exemplars such as D4Science~\citep{assante2019enacting} and CERN VRE~\citep{gazzarrini2024virtual} are often implemented as an integrated environment, incl., a catalog of research assets, a workflow management system, a data management layer, and a bunch of tools that enable user collaboration. Both CERN VRE and D4Science integrate the Jupyter for documenting and recording analytics processes. The D4Science computing infrastructure is geographically distributed across four main sites~\citep{Assante2023} and managed across different administrative domains. 
Since D4Science relies on a single entity to store and manage user data, it should be attributed to centralized identity management. 

Google Colab is a hosted Jupyter Notebook service that provides free-of-charge access to computing resources, including Graphics Processing Units (GPUs) and Tensor Processing Units (TPUs), especially well suited to machine learning, data science, and education\footnote{\url{https://colab.research.google.com/}}. It has a built-in mechanism in the Web-based interface to copy or share a Notebook. Since the computing resources and the OAuth service for authentication are mainly from Google, we place its identity management in the centralized zone. Colab shares some features with D4Science about the sharing policy definition, for instance, both can manage access control by defining who has what access for which resource. Based on this, we pose them in the decentralized management zone regarding the sharing policy aspect.  

WorkflowHub~\citep{goble2021implementing}, evolved by myExperiment\footnote{\url{https://www.myexperiment.org/about}}, is a collaborative environment where scientists can safely publish their workflows and in silico experiments, share them with groups and find those of others~\cite{de2008myexperiment}. It is a domain-agnostic workflow registry\footnote{\url{https://workflowhub.eu/}} designed around FAIR (Findable, Accessible, Interoperable, and Reusable) principles. WorkflowHub brings together scientists from different countries and across individual research institutes, to contribute to the federated registry. WorkflowHub and JupyterHub share some features that allow users can organize space, teams, and their workflows; yet, managing JupyterHub on a cluster is more centralized than WorkflowHub, in terms of the workflow orchestration and user identities. Since the main development, hosting, and maintenance of the WorkflowHub is done by a single institute, we see that identity, workflow orchestration, and asset transfers are still in centralized cooperation mode. However, it provides different levels of sharing permissions for users to define access control to various research assets within or across teams or organizations. Hence, the activities regarding sharing policies behave as decentralized.

With the advent of the infrastructure of Web3, decentralized cooperation solutions have emerged for data-centric scientific research, that aims to address the shortcomings of the current centralized solutions. Integrating decentralized technologies, such as blockchains, smart contracts, decentralized storage and protocols, and Web3 wallet for research data management marks a significant shift towards decentralized data governance and autonomy over research lifecycle management~\citep{blythman2022libraries, warnat2021swarm, ding2022desci}. Cao~\citep{cao2022decentralized} describes a vision of decentralized intelligence systems and services leveraging intelligence at edge, blockchains, Metaverse, Web3, and DeSci. Blythman \textit{et al.}~\cite{blythman2022libraries} explore Web3 wallets, Peer-to-Peer (P2P) marketplaces, decentralized storage and compute, and decentralized autonomous organizations to address the limits of the centralized management, decisions, and governance for DeAI. 

Although these visions and approaches are close to the decentralized capabilities in our work; there is a big difference between their works and ours. This work, based on our prior work NaaVRE~\citep{zhao2022notebook}, aims to bridge the gap between Jupyter-enabled VREs and decentralized cooperation modes with identity management, asset sharing and custom policy definitions, and decentralized workflows (See in Figure~\ref{fig:class}). 


\section{Decentralized Virtual Research Environment}\label{sec:D-VRE}
This section illustrates the system design and technical architecture of the D-VRE that bridges the research gaps identified in Section~\ref{sec:related}.

\subsection{Conceptual Framework for D-VRE} \label{sec:ConceptualFramework}
From the foundation of the embedded VRE solution using Jupyter~\citep{wang2022scaling, zhao2022notebook}, we have an attempt to create a decentralized collaborative environment in the community by embedding essential D-VRE components and corresponding logical layers. Figure~\ref{fig:arch} provides a high-level overview of the D-VRE conceptual framework, starting from the Jupyter-enabled private research environment. 
\begin{figure*}[!htb]
    \centering
    \includegraphics[width=\linewidth]{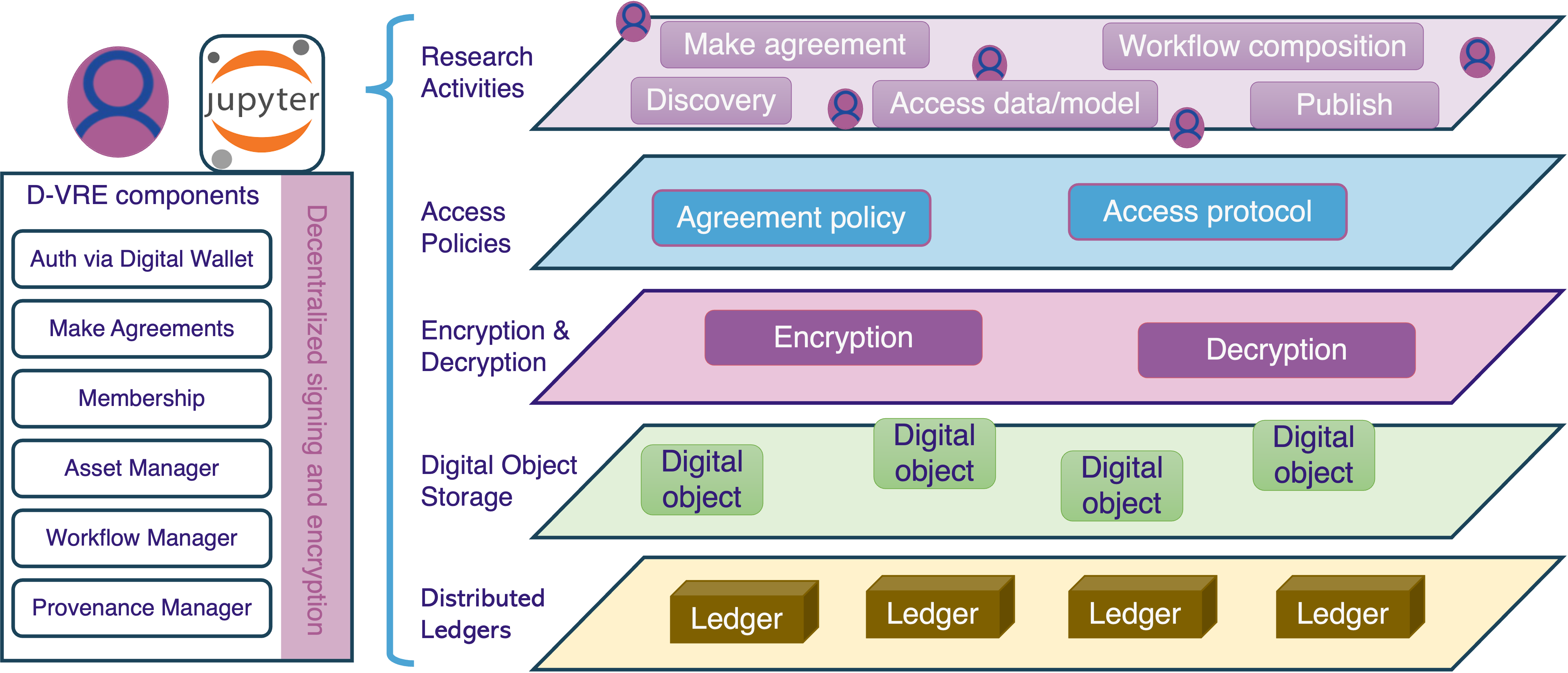}
    \caption{An overview of the D-VRE conceptual framework with the key components and logical layers.}
    \label{fig:arch}
\end{figure*}

\subsubsection{Jupyter-enabled Private Research Environment}
A research environment can be a collection of research assets, storage and computing, and software tools supporting research activities and managing research assets within the lifecycle context, e.g., the database, e-notebooks, repositories, local storage and computing, and private cloud resources. In the context of the research lifecycle, this space is usually responsible for independent research and creating knowledge within a private zone. 
This work has chosen JupyterLab as the interactive computing environment rooted in the private research environment, because it allows users to customize the behavior of the various pages for designing D-VRE components. Such a design allows users to seamlessly access D-VRE features through Graphical User Interfaces (GUIs) and services without an extra learning curve while reducing the effort. It can also save storage and computing costs by avoiding creating many comprehensive VREs. In other words, if the existing research platforms on HPC or Clouds can use JupyterLab as the companion application, they will also be able to seamlessly install and access to D-VRE components to activate the blockchain capabilities instead of reinventing the wheel.

\subsubsection{D-VRE Components}
We propose the following essential components for enabling decentralized identity, custom sharing policy definition, secure data transfer or sharing, and decentralized workflow management in the Jupyter-enabled private research environment. 

\vspace{2mm}
\noindent
{\textbf{Auth via Digital Wallet}}
shall ensure the decentralized identity function for distributed users to engage into the D-VRE network via individual digital wallets. It allows users and entities to join the D-VRE network effortlessly, enabling interaction with online services as authenticated users without a centralized oversight body, which emphasizes accessible yet secure engagement in the blockchain networks. 

\vspace{2mm}
\noindent
{\textbf{Make Agreements},} 
as one of D-VRE's most novel features, shall empower end-users to forge and customize the collaboration agreements, ensuring that precise, mutually agreed-upon terms govern collaborations while automating smart contracts. This aspect is vital for delineating the scope and rules of engagement within collaborative research.

\vspace{2mm}
\noindent
{\textbf{Membership}}
shall allow a collaboration initiator or creator to assign several users to become group members to access the research assets under the defined access conditions quickly. This component facilitates the efficient management of virtual research groups in the D-VRE network. 

\vspace{2mm}
\noindent
{\textbf{Asset Manager}}
is the component that streamlines the administration of research assets for end-users. This function shall facilitate users in managing the assets with diverse groups while catering to both shared and private data.

\vspace{2mm}
\noindent
{\textbf{Workflow Manager,}}
coupled with the Asset Manager, is the module that enables collaborative workflow management. Our use cases include federated learning workflows with decentralized data stored on heterogeneous infrastructures~\citep{kontomaris2023cwl} and scheduling privacy-aware workflows over hybrid Clouds~\citep{wang2024price}. This function shall be integral to the collaborative PoC workflows in the D-VRE ecosystem and enhance the efficiency and effectiveness of research collaborations within private groups.

\vspace{2mm}
\noindent
{\textbf{Provenance Manager}}
is an essential feature that shall maintain a transparent record of any changes made to a digital research object. We consider a combined solution to this layer, for example, utilizing secure transparent records of blockchain and research metadata approaches, e.g., Research Object Crate (RO-Crate)~\citep{soiland2022packaging} and Schema.org~\citep{barker2014schema}. 


\subsubsection{Logical Layers}
In the framework, we consider the following five logical layers to achieve the above D-VRE components. 

\vspace{2mm}
\noindent
{\textbf{Research Activity Layer}} is towards individual users, organizations, and research communities. Research activities among distributed researchers in scientific communities might vary across different disciplines. In this layer, we point out several main activities related to the research collaboration --- viz discovery of resources, making agreements for sharing research assets that cover ownership, access control, and data usage rights, as well as workflow composition and publishing.

\vspace{2mm}
\noindent
{\textbf{Access Policy Layer}} aims to collect users' input for making agreements from the human-in-the-loop interactive interface and translate it into the D-VRE system's agreement policy and access protocol. This layer shall facilitate efficient agreement-making and enforce the access control policies.

\vspace{2mm}
\noindent
{\textbf{Encryption and Decryption Layer}} is a safeguard of the secure connection between the Jupyter-enabled private research environment and the D-VRE ecosystem. Cryptography is crucial for data integrity, availability, and confidentiality. We assume that decentralized signing and encryption can be embedded in the Jupyter-enabled private research environment. Thus, this layer shall introduce robust cryptographic solutions for the system with them.

\vspace{2mm}
\noindent
{\textbf{Digital Object Storage Layer.}} Considering the long-live research activities, the system shall provide robust data storage solutions and archiving capabilities for various data types, e.g., documents, project code files, datasets, AI models, and e-notebooks. In accordance with the FAIR principles, the digital object shall be stored with the data itself, a variable amount of metadata, and a globally unique identifier~\citep{schultes2019fair}, making it findable, accessible, interoperable, and reusable. 

\vspace{2mm}
\noindent
{\textbf{Distributed Ledger Layer.}} 
It is known that a distributed ledger is the consensus of shared and synchronized across various sites, institutions, or geographical locations, and accessible by multiple users. Any changes or additions made to the ledger are instantly or within minutes reflected and copied to all participants~\citep{rauchs2018distributed}. The most famous one is the blockchain network. This layer shall support the underlying data management with such features. 

\begin{figure*}[!htb]
    \centering
    \includegraphics[width=\linewidth]{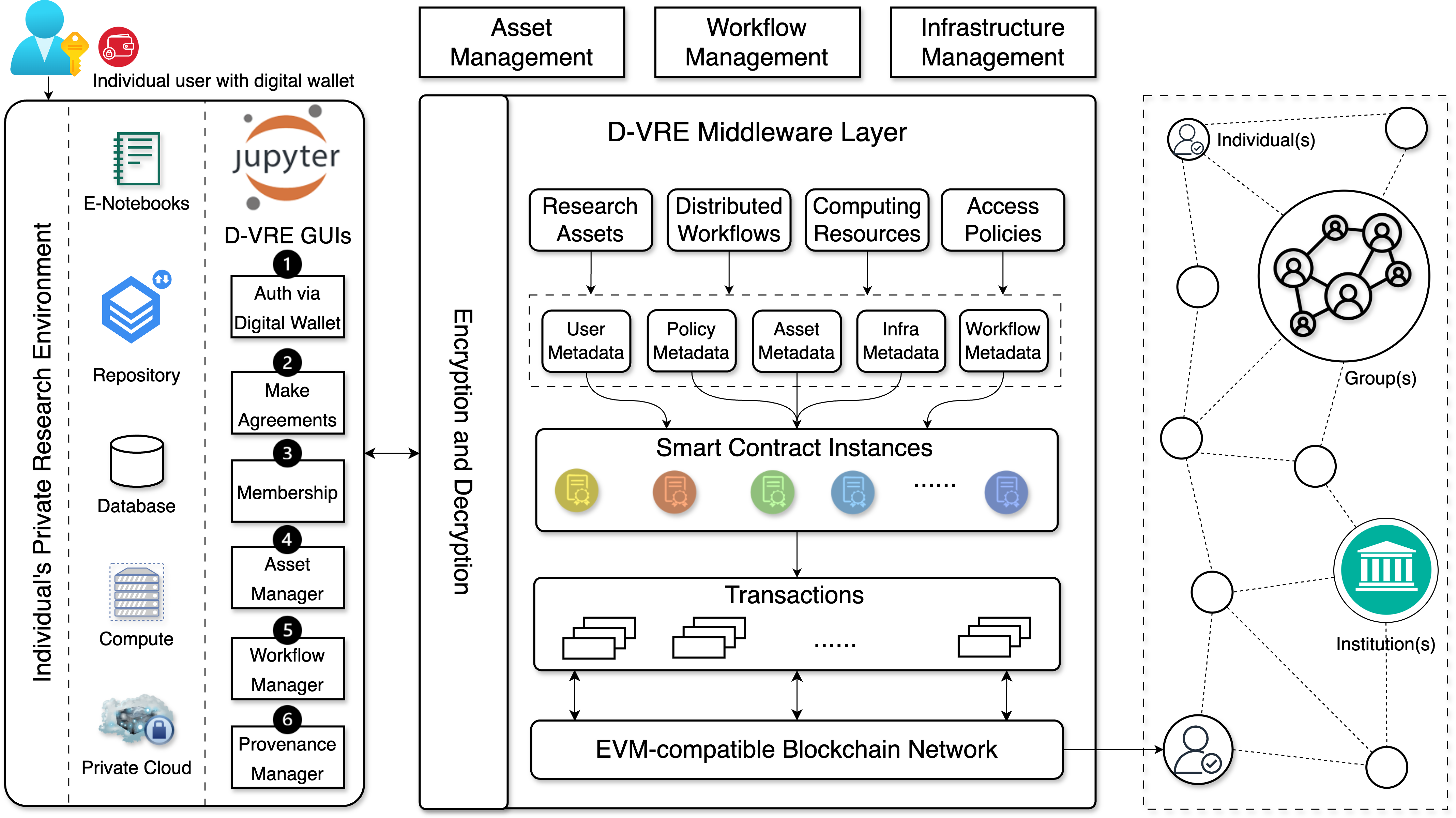}
    \caption{An overview of the D-VRE architecture, in conjunction with the individual's private research environment (left), D-VRE middleware layer and the management of asset, workflow, and infrastructure above (middle), and different potential customers in the blockchain network (right).}
    \label{fig:workflowArchitecture}
\end{figure*}

\subsection{D-VRE Architecture}
Based on the conceptual framework design, we propose the D-VRE system architecture, as illustrated in Figure~\ref{fig:workflowArchitecture}. It visually represents the overview of the D-VRE, and highlights the modules within the D-VRE middleware and their interactions between the individual's private research environment and the decentralized network hosting different customers. Although users' private research environments may vary in the great research picture, the D-VRE middleware shall be able to adapt to various environments via Jupyter. Thus, individual users, institutes, or groups using Jupyter can easily engage in the community. 

In our design, the metadata of research objects, including research assets, distributed workflows, computing resources, and access policies, shall be published and stored on the blockchain, e.g., the Ethereum blockchain. An Ethereum account could work across public networks (e.g., Ethereum Mainnet and Testnets) and private networks (e.g., development and consortium networks). Stakeholders can choose a preferable operational model to construct blockchain networks based on their requirements and budgets. Leveraging blockchain, the system can verify the processes and automatically deploy a number of immutable smart contract instances in the blockchain network. It thus ensures that all participants adhere to the agreed-upon terms. Consequently, it can pave a powerful way to enhance the integrity, traceability, and security of the digital objects.

Regarding the application scenarios, we refer to the prior research experience and user requirements from several EU projects such as CLARIFY\footnote{\url{http://www.clarify-project.eu/}}, BlueCloud2026\footnote{\url{https://blue-cloud.org/about-blue-cloud-2026}}, ENVRI-Hub NEXT\footnote{\url{https://envri-hub.envri.eu/}}, and LifeWatch ERIC\footnote{ \url{https://www.lifewatch.eu/}}. In general, there are three main categories regarding collaboration and sharing: (1) research asset sharing, (2) collaborative workflows in machine learning and data science, and (3) federated infrastructures. This paper focuses on research asset sharing. Regarding collaborative workflows with D-VRE, we have considered potential applications for data quality control (ocean data and medical images), digital twin applications, and distributed machine learning workflows (at the training phase and inference phase, respectively). Furthermore, participants in the D-VRE might want to share idle storage and computing resources. From the perspective of federated infrastructure, examples include outsourcing computation tasks to remote infrastructure in multi-institute collaborations.

\subsection{Technology Choices}\label{sec:tech}
After the analysis of the literature survey and technology investigation, we select the following key technologies. 

\vspace{2mm}
\noindent 
{\textbf{Web Technology}} is critical for developing Web-based applications. It can be utilized to design and implement the user interface and interactions that present information to users, capture user input, and access functions and services. Our D-VRE shall involve both Web 2.0 and Web 3.0 technologies. To develop D-VRE GUIs, we have considered common Web 2.0 technologies such as HTML, CSS, and JavaScript (JS), including Node.js, Express.js, Vue.js, and React. We investigate Node.js and React for the webpage design. Regarding Web 3.0, we have investigated the following key technologies for further implementation. 


\vspace{2mm}
\noindent 
{\textbf{Blockchains}} are the first citizens of Web 3.0. We look at Ethereum Virtual Machine (EVM)-compatible blockchains (e.g., Ethereum\footnote{ Ethereum. \url{https://ethereum.org/}}, Polygon\footnote{Polygon. \url{https://polygon.technology/}}, Avalanche\footnote{Avalanche. \url{https://www.avax.network/}}) and non-EVM blockchains (e.g., Hyperledger Fabric\footnote{Hyperledger Fabric. \url{https://www.hyperledger.org/projects/fabric}} and Cardano\footnote{Cardano. \url{https://cardano.org/}}). We have chosen Ethereum in this work because of the potential of full decentralization (e.g., public blockchain networks), native cryptocurrency support, and a broad developer community. Ethereum provides a comprehensive development toolchain for researchers and developers to implement smart contracts, DApps, and secure transactions between nodes in the network using Web3.js and Ethers.js. This is crucial for custom agreements or sharing policies and automatically enforcing and executing contract terms when predefined conditions are met. \newline

\vspace{2mm}
\noindent 
{\textbf{Digital Wallet,}} which injects Web3 objects into the webpage's JavaScript content, enables D-VRE components to interact with the decentralized Web, via individual user's browsers (e.g., Chrome, Firefox, and Safari). We have chosen Metamask~\citep{lee2019using}, a self-custody digital wallet, for the first component implementation. MetaMask can act as the user's digital identity and bank. It allows users to sign transactions and messages without revealing their private keys and offers a secure way to prove ownership and authorize processes.

\vspace{2mm}
\noindent
{\textbf{Lit Protocol.}} To enhance the distributed trust and security of the system, we also investigate advanced security mechanisms for encryption and decryption layer. It is well-known that MetaMask as a self-custody wallet might face the issue of a single point of failure. To address this, we have chosen Lit protocol, as the key management modular and enhance the secure data sharing with access conditions. It distributes encrypted key shares across the Lit network by harnessing multi-party computation, threshold secret schemes, and trusted execution environments to ensure that node operators never have access to key shares or computations processed within the nodes. Lit allows users to manage sovereign identities on the open Web with self-defined flexible access control\footnote{Lit protocol. \url{https://www.litprotocol.com/}}, ensuring robust, efficient, and secure DApp solutions embedded in the system. 


\vspace{2mm}
\noindent
{\textbf{IPFS,}} a.k.a., InterPlanetary File System~\citep{benet2014ipfs} can provide a set of composable, P2P protocols for addressing, routing, and transferring content-addressed data in a decentralized file system\footnote{IPFS. \url{https://docs.ipfs.tech/}}. The main reasons why we choose IPFS include (1) it can lower the cost associated with data storage on the blockchain, (2) the adoption of IPFS enhances data availability and persistence, and (3) it aligns with the open-source and academic communities' preference for cost-effective decentralized data management solutions.

\section{Prototype, Experiments, and Evaluation}\label{sec:prototype}
This section illustrates the current implementation of the D-VRE prototype, details the experimental study, and demonstrates the visualized results with a case study and its validation. 

\begin{figure*}[!htb]
    \centering
    \begin{lstlisting}
    
    contract GroupContract {
        struct ContractDetails {
            string groupName;
            address groupOwnerAddress;
            string permissions;
            string[] organizations;
            string[] countries;
        }
    ...
        modifier onlyGroupOwner() {
            require(msg.sender == contractDetails.groupOwnerAddress, "Only group owner can call this function");
            _;
        }
        function addFilesToGroup(FileDetails[] memory fileDetails) public {
            for (uint256 i = 0; i < fileDetails.length; i++) {
                FileDetails memory fDetail = fileDetails[i];
                string memory IPFSHash = fDetail.IPFSHash;
                sharedIPFSHashes[IPFSHash] = true;
                addedFileDetails.push(fDetail);
                emit Success("Files successfullly shared in the group");
            }
        }
        function associateUsersToGroup(UserInput[] memory users) public onlyGroupOwner{
            for (uint i = 0; i < users.length; i++) {
                setUserAccess(users[i].eoaAddress, users[i].accessFrom, users[i].accessTo);
            }
            emit Success("Users successfully added to the group");
        }
    }
    \end{lstlisting}
    
    \caption{Example of the \texttt{\textbf{GroupContract}} smart contract.}\label{fig:smart contract}
\end{figure*}


\subsection{Prototype Implementation}\label{sec:RE}

We have configured INFURA\footnote{INFURA. \url{https://docs.infura.io/api}} and Pinata\footnote{Pinata. \url{https://docs.pinata.cloud/api-reference/introduction}} API keys to simplify the management of blockchain infrastructure and IPFS. We implemented smart contracts in Solidity 0.8.0 and developed custom Web3 applications and extensions for D-VRE, which are seamlessly integrated into the JupyterLab environment. To evaluate the D-VRE prototype in a realistic setting, we deployed these smart contracts on one of the Ethereum Testnets --- viz the Sepolia network. Moreover, it is worth noting that the demonstration and case study results presented here are, in whole or in part, based on the collaborative research activities in the CLARIFY project. 

\subsubsection{D-VRE Smart Contracts}
In this prototype, we implemented four smart contracts, namely \texttt{\textbf{UserMetadataFactory}}, \texttt{\textbf{UserMetadata}}, \texttt{\textbf{PolicyManager}}, and \texttt{\textbf{GroupContract}}, by leveraging the idea of human-in-the-loop smart contracts. The first two contracts were designed for user identity management, for example, user registration or login, authentication, and authorization. And the user metadata factory is the parent contract of the user metadata contract. Similarly, the group contract is the child contract of the policy manager. Figure~\ref{fig:smart contract} is an example of Solidity implementation of the \texttt{\textbf{GroupContract}}. Any user with an Ethereum Externally Owned Account (EOA) can create private groups to become a group owner. In the \texttt{\textbf{ContractDetails}} struct, it contains the group name, group owner's address, permissions, organizations, and countries, as well as a modifier, and two functions in the listings. The group owner can invite other users to become group members via the function \texttt{\textbf{associateUsersToGroup}}. Participants in the group can share digital objects via IPFSHash to reduce the storage cost in the blockchain through the \texttt{\textbf{addFilesToGroup}} function. Note that we also implement the access duration control as some shared data might be timeliness within a research lifecycle context. 

\subsubsection{Lit-based Encryption and Decryption}
IPFS uses transport encryption but not content encryption, if anyone has the content identifier (IPFSHash code) of a digital object, then he or she can download and view that data via the IPFS gateway. To address this issue, we leverage Lit to encrypt the content stored in IPFS and guarantee that only registered users can decrypt it with qualified access control conditions. 

\input{lit-encrypt}

\input{lit-decrypt}

The main procedures for encrypting files and uploading them to IPFS have been written in the algorithm pseudo-code~\ref{alg:lit-encrypt}. The user can create an access control condition and combine it with selected files, e.g., stored in IPFS. The user holds the encryption metadata, i.e., access control condition, IPFSHash, the time window of the private data, and a ciphertext. This mechanism guarantees that only authorized users can successfully decrypt the ciphertext and access the shared data while reducing the data storage cost in the blockchain. More decryption details have been presented in Algorithm~\ref{alg:lit-decrypt}. 

For more complex security or access control requirements, we can customize the access control condition definitions. For example, the security module be adopted in various application scenarios by setting the parameters, such as blockchain network or chain name, lit node network name, and so forth. 


\begin{figure*}[!htb]
    \centering
    \includegraphics[width=\linewidth]{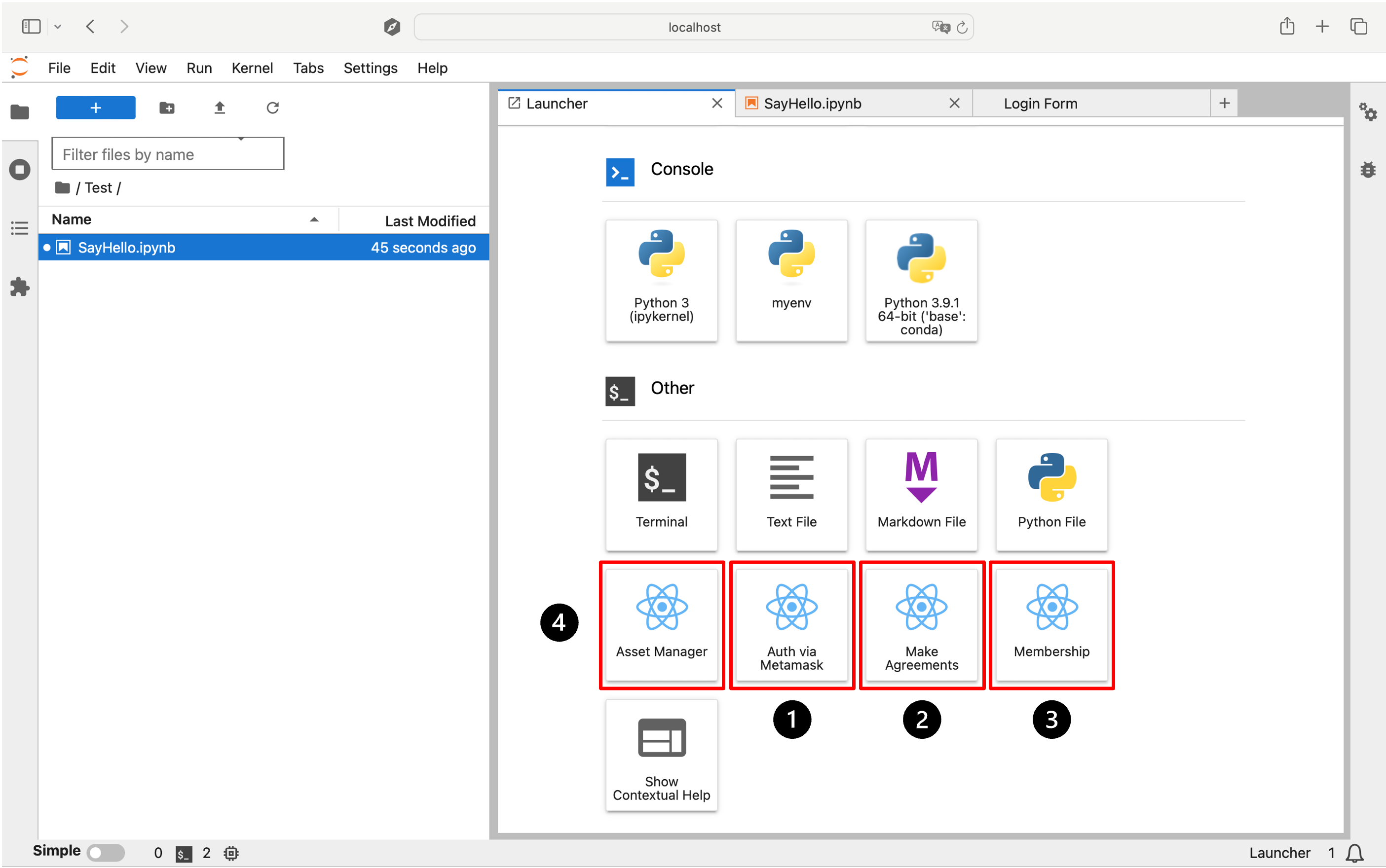}
    \caption{The screenshot of the embedded D-VRE components in the JupyterLab environment: \ding{182} Auth via MetaMask, \ding{183} Make Agreements, \ding{184} Membership, \ding{185} Asset Manager.} \label{fig:interfaces}
\end{figure*}

\begin{figure*}[!hbt]
    \centering
    \includegraphics[width=\linewidth]{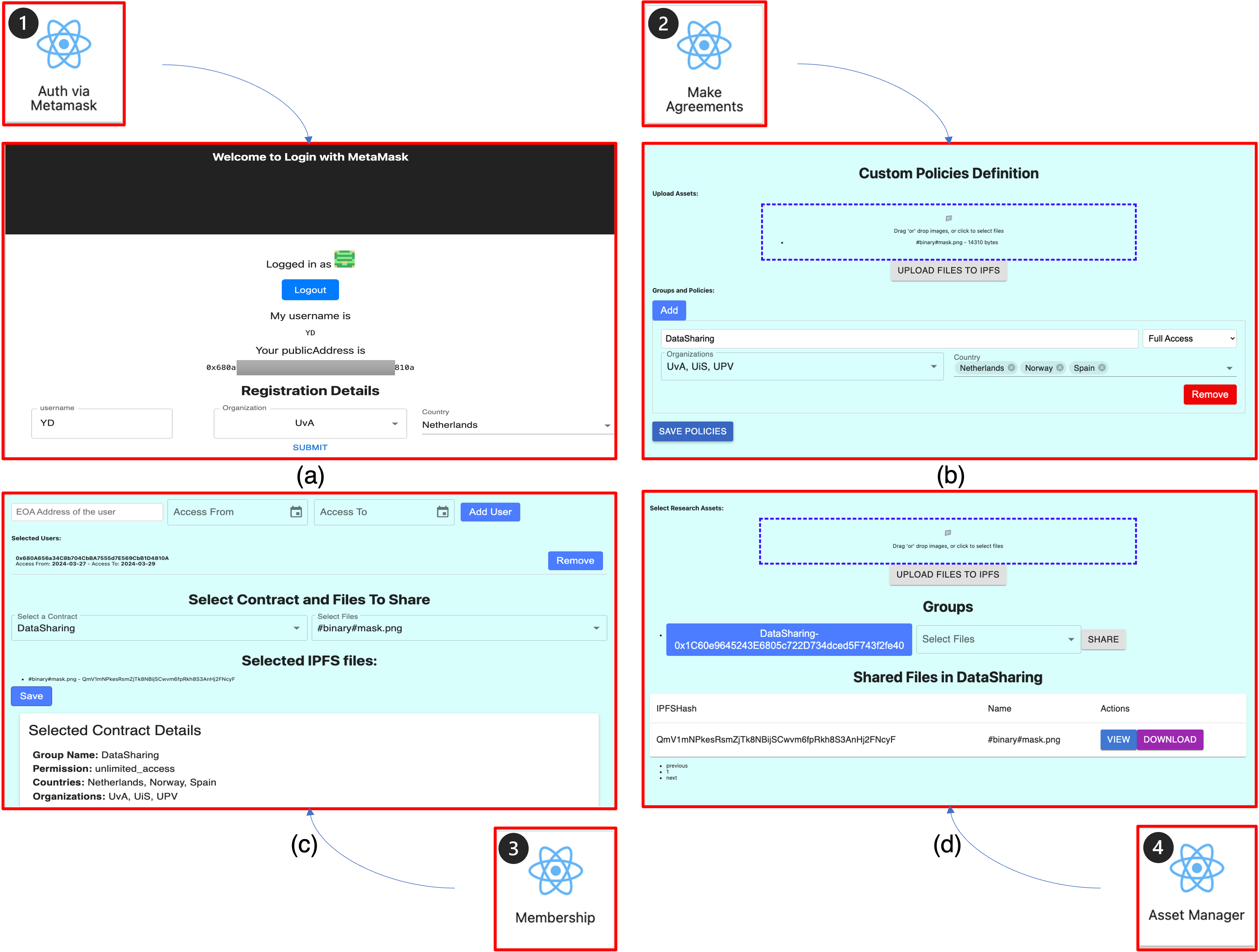}
    \caption{An overview of the collection of the D-VRE GUIs: (a) Auth via MetaMask, (b)Make Agreements, (c) Membership, and (d) Asset Manager.}
    \label{fig:dvre}
\end{figure*}

\subsubsection{Jupyter Integration}
Figure~\ref{fig:interfaces} presents the screenshot of our current D-VRE prototype implementation embedded in JupyterLab. Four out of six (4/6) essential D-VRE modules have been seamlessly integrated into the launcher of the JupyterLab environment: \ding{182} Auth via MetaMask, \ding{183} Make Agreements, \ding{184} Membership, \ding{185} Asset Manager. In Figure~\ref{fig:dvre}, we present a collection of critical GUIs from these four components. 
Note that the rest features of D-VRE, such as Workflow Manager and Provenance Manager, are still under development and work in progress.

\subsubsection{Auth via MetaMask}
Figure~\ref{fig:dvre}(a) presents the screenshot of the login page in the \ding{182} Auth via MetaMask component, where Jupyter users can register as D-VRE users and login to the D-VRE system via Metamask. Specifically, each user can register with the MetaMask's \texttt{\textbf{publicAddress}} (e.g., \texttt{0x680Axxxx810A}), the \texttt{\textbf{username}}, \texttt{\textbf{organization}}, and \texttt{\textbf{country}} details. When interacting with D-VRE, the user's MetaMask public address is by default associated with the D-VRE account for decentralized signing and encryption.

\subsubsection{Make Agreements}
In the \ding{183} Make Agreement component, we have implemented flexible group creation with custom terms of agreement. 
Figure~\ref{fig:dvre}(b) presents the details of this feature, including custom policies of data sharing, uploading assets (files) to IPFS, and group creation details. 
The user can first upload an asset, for example, an image titled \texttt{\#binary\#mask.png} from the local system to IPFS. In the PoC implementation, the IPFS storage was configured with the free-of-charge account of the Pinata IPFS gateway, which freely supports a total of 500 pinned files and 1 GB of storage per workplace member. Then, the user can add one or multiple groups for data sharing purposes by customizing the name --- \texttt{\textbf{DataSharing}}, permission settings --- \texttt{\textbf{Full Access}}, cross-organizational collaboration among \texttt{\textbf{UvA}}, \texttt{\textbf{UiS}}, and \texttt{\textbf{UPV}}, and countries --- \texttt{\textbf{Netherlands}}, \texttt{\textbf{Norway}}, and \texttt{\textbf{Spain}}. The user saves the input by clicking on the button \texttt{\textbf{SAVE POLICIES}}. Finally, the system will automatically generate a group smart contract instance, written into code and deploy it to the Ethereum network.

\subsubsection{Membership}
As shown in Figure~\ref{fig:dvre}(c), a group owner can increase group members by adding the EOA address of the users, which refers to the user's \texttt{\textbf{publicAddress}} introduced in the Metamask. Meanwhile, the user can customize the access period of the group and shared research assets inside. For example, the user (i.e., \texttt{0x680Axxxx810A}) was selected by the \texttt{\textbf{DataSharing}} group owner and granted unlimited access to the shared data, available from \texttt{\textbf{2024-03-27}} to \texttt{\textbf{2024-03-29}} in the \ding{184} Membership page. 

\subsubsection{Asset Manager}
Figure~\ref{fig:dvre}(d) showcases the webpage of an authenticated user in the \ding{185} Asset Manager interface. It lists the group and the shared files in the group. The authenticated user or group member (e.g., \texttt{0x680Axxxx810A}) can select local files to upload to IPFS for group sharing. In addition, the user can view and download the shared file \texttt{\#binary\#mask.png}. If a user has become a group member and has been granted access permission, he will see all the involved groups and shared files through the action buttons \texttt{\textbf{VIEW}} and \texttt{\textbf{DOWNLOAD}}.

\subsection{Experimental Study}
To simplify the analysis of gas usage of the smart contract implementations in our D-VRE, we conducted an experimental study by deploying the implemented smart contracts on the Ganache\footnote{Ganache. \url{https://github.com/trufflesuite/ganache}}. As an essential tool in the Ethereum development ecosystem, it provides a personal blockchain that developers can use to deploy contracts, run tests, and execute commands without the need for a live network. We leverage the retrieved first account from the local Ganache node to execute the interfaces and estimate different types of gas according to the D-VRE smart contract implementations. Figures~\ref{fig:deploy} and~\ref{fig:functions} present the results of this experimental study. 

\begin{figure}[!htb]
    \centering
    \includegraphics[width=\linewidth]{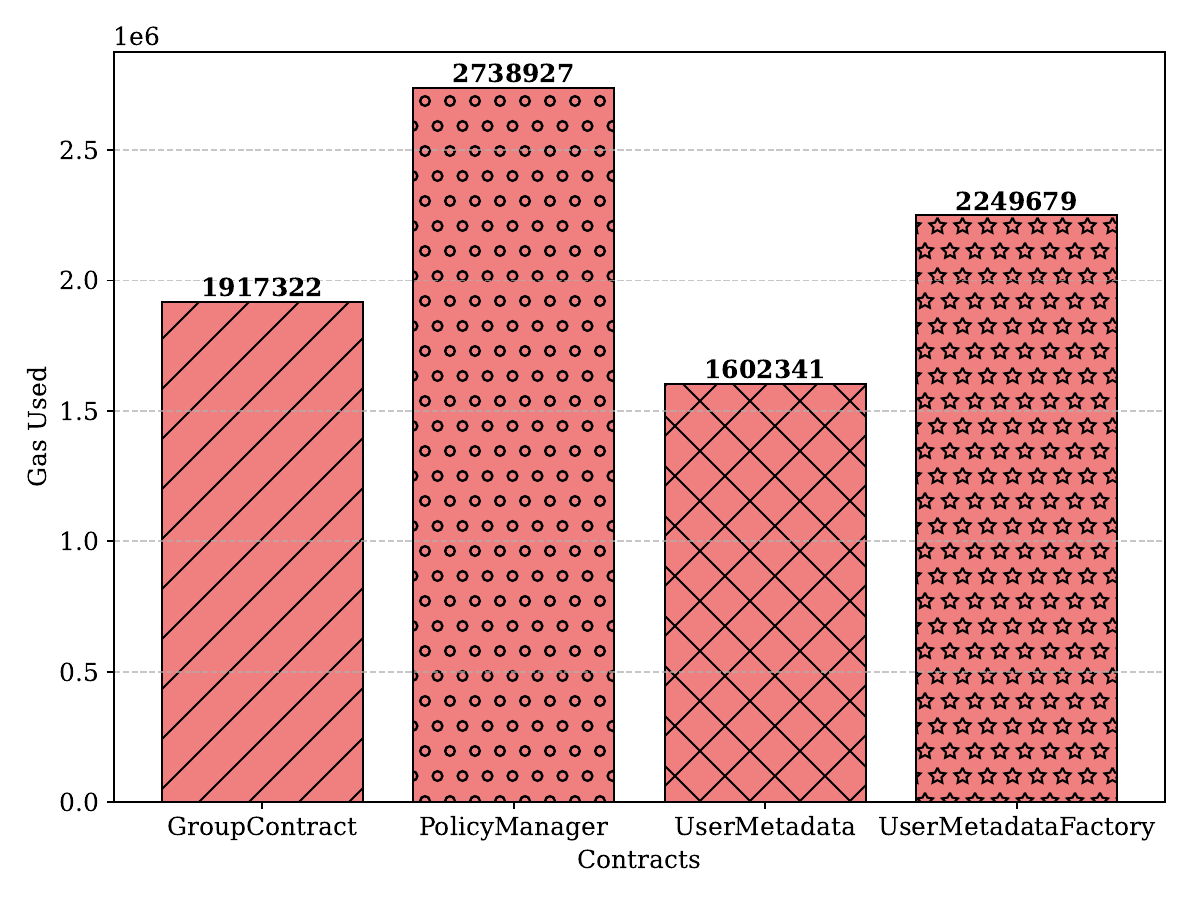}
    \caption{An overview of the gas usage for the D-VRE smart contract deployment.}
    \label{fig:deploy}
\end{figure}
\begin{figure*}[!htb]
    \centering
    \includegraphics[width=\linewidth]{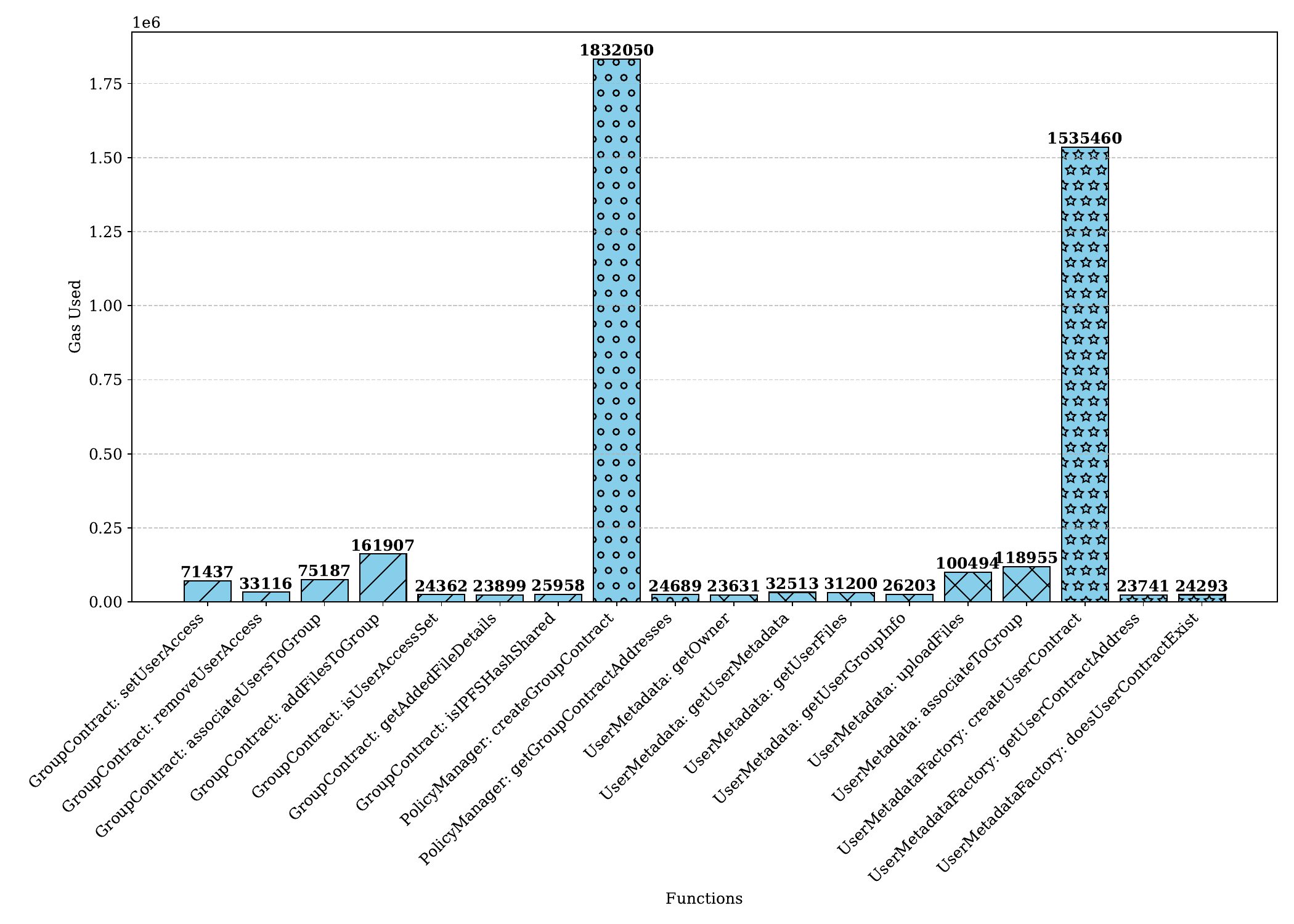}
    \caption{An overview of the gas consumption of all functions in smart contracts.}
    \label{fig:functions}
\end{figure*}

We compare the gas units consumed by the deployment of different smart contracts, as presented in Figure~\ref{fig:deploy}. It is obvious that deploying parent contracts i.e., \texttt{\textbf{PolicyManager}} and \texttt{\textbf{UserMetaFactory}}, consumes more than child smart contracts, i.e., \texttt{\textbf{GroupContract}} and \texttt{\textbf{UserMetadata}}, respectively. The first place (\texttt{PolicyManager}, 2,738,927 gas used) has 498,248 more than the second one (\texttt{UserMetaFactory}, 2,249,679 gas used). In contrast, \texttt{\textbf{UserMetadata}} contract costs the lowest (1,602,341 gas units), less than the \texttt{\textbf{GroupContract}} (1,917,322 gas usage), and their difference is 314,981. 
Figure~\ref{fig:functions} compares the gas used for all functionalities in each smart contract. It is noticeable that the function \texttt{\textbf{createGroupContract}} from the \texttt{\textbf{PolicyManager}} smart contract used the most gas. Similarly, the function \texttt{\textbf{createUserContract}} places the second. Notably, the gas costs for creating a group and user metadata contracts are estimated at 1,832,050 and 1,535,460, respectively, far above (nearly 7 and 6 times) the used gas of other functions. 

These differences in gas consumption highlight the operations' efficiency and resource use of the current implementations. There is still optimization space to reduce gas consumption by optimizing interface design and implementation in smart contracts.

\subsection{Feedback from the Case Study}
Analysis of the researchers' feedback is beneficial for the usability, ease of use, impact, and improvement suggestions of the D-VRE prototype. We disseminated our D-VRE software prototype and surveyed the CLARIFY community by asking the following four questions. 

\textbf{$Q_1:$} Do you think the D-VRE features are functional or helpful in the research lifecycle? And why?

\textbf{$Q_2:$} Are the D-VRE features easy to use? And why?

\textbf{$Q_3:$} Will you be interested in using the proposed D-VRE in your daily research practices, for example, in building decentralized collaboration and data-sharing? 

\textbf{$Q_4:$} (optional) Do you have any suggestions for improvement for a research collaboration scenario you'd like to facilitate during the D-VRE dissemination?

\noindent
\paragraph{\textbf{Usability.}}
We collected five responses to $Q_1$ about usability and usefulness, and 3/5 of the participants gave very positive feedback. They like the idea of defining policies by individual researchers, making data available only during a specific period, and the encryption and decryption feature for sensitive data. The rest of the participants think the D-VRE features may be useful since it still looks a bit limited: 
\begin{itemize}
    \item ``The D-VRE looks like an institute-level feature not so functional for individual researchers. Maybe organization design helps.'' 
    \item ``It is unclear how much leverage we get using D-VRE instead of the simpler sharing option on Google Colab. Maybe comparing existing solutions can highlight their usefulness in different cases.''
\end{itemize}
Note that we have compared existing solutions, incl., Google Colab, in Section~\ref{sec:related}.

\noindent
\paragraph{\textbf{Ease of Use.}} Based on the responses to $Q_2$, all participants think the D-VRE features are easy to use because (1) the GUI looks simple and understandable, (2) buttons and features are clearly named and straight-to-the-point, and (3) the video tutorials are well detailed. However, we also received some suggestions about GUI improvements:  
\begin{itemize}
    \item ``Hopefully a better visualization in GUIs.'' 
    \item ``Maybe A/B testing with different features can help understand how different groups can find it easier without knowing much about notebooks.''
\end{itemize}

\noindent
\paragraph{\textbf{Impact.}}
Regarding whether they will be interested in using the proposed D-VRE in their daily research practices, we received 4/5 positive feedback about $Q_3$. 
One participant commented that it needs to be robust, and he or she wanted to know if the D-VRE supports large-scale datasets and training. The biggest concern in the responses is the volume of data. Participants wonder how much it can handle. 


\section{Discussion and Open Challenges}\label{sec:discussion}
This section discusses the limitations of the current D-VRE, open challenges, and possible improvements in future directions.

D-VRE, based on a decentralized future, is expected to impact many scientific domains that rely on research data and models from distributed sources, using Jupyter for data analytics, and large-scale computation in the age of Web 3.0 --- give power back to the users in the form of ownership. We argue that D-VRE can serve as a novel framework for achieving a decentralized virtual research environment ecosystem and our vision is to make it a working system that can serve as a key driver for decentralized science. To achieve that goal, it must be researched, developed, tested, deployed, and evaluated based on the infrastructure of Web 3.0 and address core research questions in scientific communities. 

The primary objective of this study was to bridge the gap between an individual's private research platform and the decentralized collaborative research ecosystem from three perspectives: asset management, workflow management, and infrastructure management. We prototyped D-VRE with seamless integration of JupyterLab. Despite this significant bridge, this study has several limitations. 

\subsection{Operation Model}
First but foremost, there is still optimization space in the operation model. Currently, the D-VRE is deployed on the Sepolia Testnet, which features a Proof-of-Stake (PoS) consensus mechanism to mimic the Ethereum Mainnet, where everyone can test running a validator. However, it might face an issue with the cheating validators or other risks related to cheating behaviors. A consortium network may be preferable because a pre-defined set of trusted nodes controls the consensus process. For instance, a group of academic institutes like universities can construct a private network, and each institute governs a single node. As a result, the institute can verify individual user identities and make research activities more reliable. In this case, it can also combine with verifiable credential models to enhance the decentralized identity management for decentralized authentication.

\subsection{Energy Consumption and Sustainability}
According to the statistics of Crypto Carbon Ratings Institute (CCRI)\footnote{\url{https://carbon-ratings.com/}, retrieved on 19 June 2024.}, the annualized electricity consumption of the Ethereum PoS is estimated approximately 0.0060 TWh across the entire global network, which corresponds to the estimated yearly carbon emissions of 0.0019MtCO2e. In contrast, the annual global data centre electricity consumption in 2022 was 240-340TWh, according to the tracking from the International Energy Agency (IEA)\footnote{\url{https://www.iea.org/energy-system/buildings\#tracking}, retrieved on 19 June 2024.}. Thus, we can say that Ethereum is green blockchain thanks to the PoS consensus mechanism. However, there is still optimization space to reduce the energy consumption by optimizing the consensus algorithms and/ or the resource utility of distributed infrastructures. For example, there might be a PoX (X is an unknown factor) consensus mechanism in the future that will make the Ethereum blockchain greener than using PoS. 

Apart from energy consumption and CO2 emissions, smart contracts or transaction optimization is another aspect to facilitate sustainability. It is known that every operation that performs a transaction or smart contract on the Ethereum platform costs certain units of gas, and the gas used denotes the metering unit for the use of the Ethereum virtual machines in the world computer. Assume that we could decrease the gas used in smart contract deployment and operations in the smart contract functionalities, it will also reduce the use of the Ethereum world computer. As such, the corresponding electricity energy and carbon emissions will be reduced. 

\subsection{Community Support}
Last but not least, our D-VRE needs the support from the community. However, one of the biggest challenges to the sustainability of any scientific tools or software is community support. There is a need to continue research and development and engage in D-VRE dissemination, including a well-documented manual, demonstration of D-VRE features, and tutorials for the target communities. 

The current prototype implementation may restrict the generalizability of our findings since it only has been disseminated in one community, i.e., CLARIFY, that shows significant potential for use in medical research, especially in supporting sensitive data sharing. The need for D-VRE to adapt to the varied demands of different communities highlights the importance of incorporating flexibility into its design. 

\section{Conclusion and Future Work}\label{sec:conclusion}
This article presents D-VRE, a significant bridge between the Jupyter environment and a decentralized ecosystem for scientific research. The PoC prototype achieved seamless integration with blockchain-empowered D-VRE applications and extensions, such as Auth via MetaMask, Make Agreements, Membership, and Asset Manager into JupyterLab. We conducted an experimental study to test all functionalities written in D-VRE smart contracts and their gas consumption. In addition, the conducted case study demonstrates the D-VRE's usability, ease of use, and potential in P2P trustworthy data sharing and collaboration scenarios. The promising results from the D-VRE prototype bridge the gap between traditional Jupyter-related virtual research environments and blockchain capabilities, paving the way for broader collaborative applications in real-world situations using Jupyter. 

In our future work, we will first complete the rest of the D-VRE components such as workflow manager and provenance manager. Then we will investigate the underlying metadata management to enhance the FAIRification in the context of research lifecycles. Moreover, we will continuously improve our D-VRE by engaging in more data-centric scientific communities and exploring more real-world use cases.

\section*{Acknowledgments}

This research was made possible through partial funding from several European Union projects: CLARIFY (860627), ENVRI-Hub Next (101131141), EVERSE (101129744), BlueCloud-2026 (101094227), OSCARS (101129751), LifeWatch ERIC, BioDT (101057437, through LifeWatch ERIC), and Dutch NWO LTER-LIFE project. 

\section*{CRediT authorship contribution statement}
\textbf{Yuandou Wang:} Conceptualization, Methodology, Software (testing and revision of existing code components), Investigation, Validation, Formal analysis, Writing – Original draft, Writing – Review \& Editing. \textbf{Sheejan Tripathi:} Methodology, Software, Validation, Formal analysis, Investigation. \textbf{Siamak Farshidi:} Writing – Review \& Editing. \textbf{Zhiming Zhao:} Conceptualization, Writing – Review \& Editing, Supervision, Project administration, Funding acquisition. 



\bibliographystyle{IEEEtran}
\bibliography{references}
\end{document}

%% file: lit-encrypt.tex
\begin{algorithm}[!htb]
\SetAlgoLined
\KwOut{a zipped folder with the encrypted file and its metadata $c$ and the response $r$.}

\KwIn{Key of a user's local storage $k$, litNetwork.}

$f, state \gets$ selected file from local storage with $k$;

client $\gets$ \textit{litNodeClient}(litNetwork);

accs $\gets$ define the access control conditions;

authSig $\gets$ check and sign on Ethereum chain; 

$c \gets$ \textit{encryptFileAndZipWithMetadata}(accs, chain, $f$, authSig, client, readme);

$r \gets$ respond to uploading $c$ and $f.name$ to IPFS gateway;

\Return $c$, $r$.\
\caption{EncryptFileAndUpload}\label{alg:lit-encrypt}
\end{algorithm}

%% file: lit-decrypt.tex
\begin{algorithm}[!htb]
\SetAlgoLined
\KwOut{a zipped folder with the decrypted file and its metadata $f$ and the response $r$.}

\KwIn{IPFS gateway, $c$, litNetwork.}

client $\gets$ \textit{litNodeClient}(litNetwork);

authSig $\gets$ check and sign on Ethereum chain; 

$f \gets$ \textit{decryptZipFileWithMetadata}($c$, authSig, client)

$r \gets$ respond to downloading $f$ with metadata from IPFS gateway;

\Return $f$, $r$.\
\caption{DecryptFileAndDownload}\label{alg:lit-decrypt}
\end{algorithm}

%% file: main.bbl
\begin{thebibliography}{10}
\providecommand{\url}[1]{#1}
\csname url@samestyle\endcsname
\providecommand{\newblock}{\relax}
\providecommand{\bibinfo}[2]{#2}
\providecommand{\BIBentrySTDinterwordspacing}{\spaceskip=0pt\relax}
\providecommand{\BIBentryALTinterwordstretchfactor}{4}
\providecommand{\BIBentryALTinterwordspacing}{\spaceskip=\fontdimen2\font plus
\BIBentryALTinterwordstretchfactor\fontdimen3\font minus \fontdimen4\font\relax}
\providecommand{\BIBforeignlanguage}[2]{{%
\expandafter\ifx\csname l@#1\endcsname\relax
\typeout{** WARNING: IEEEtran.bst: No hyphenation pattern has been}%
\typeout{** loaded for the language `#1'. Using the pattern for}%
\typeout{** the default language instead.}%
\else
\language=\csname l@#1\endcsname
\fi
#2}}
\providecommand{\BIBdecl}{\relax}
\BIBdecl

\bibitem{sidey2019machine}
J.~A. Sidey-Gibbons and C.~J. Sidey-Gibbons, ``Machine learning in medicine: a practical introduction,'' \emph{BMC medical research methodology}, vol.~19, pp. 1--18, 2019.

\bibitem{monteleoni2013climate}
C.~Monteleoni, G.~A. Schmidt, and S.~McQuade, ``Climate informatics: accelerating discovering in climate science with machine learning,'' \emph{Computing in Science \& Engineering}, vol.~15, no.~5, pp. 32--40, 2013.

\bibitem{mathur2023machine}
P.~Mathur and M.~Mathur, ``Machine learning ensemble species distribution modeling of an endangered arid land tree tecomella undulata: a global appraisal,'' \emph{Arabian Journal of Geosciences}, vol.~16, no.~2, p. 131, 2023.

\bibitem{henderson2020accelerating}
M.~L. Henderson, W.~Krinsman, S.~Cholia, R.~Thomas, and T.~Slaton, ``Accelerating experimental science using jupyter and nersc hpc,'' in \emph{Tools and Techniques for High Performance Computing: Selected Workshops, HUST, SE-HER and WIHPC, Held in Conjunction with SC 2019, Denver, CO, USA, November 17--18, 2019, Revised Selected Papers 6}.\hskip 1em plus 0.5em minus 0.4em\relax Springer, 2020, pp. 145--163.

\bibitem{perkel2018jupyter}
J.~M. Perkel, ``By jupyter, it all makes sense,'' \emph{Nature}, vol. 563, no. 7729, pp. 145--146, 2018.

\bibitem{gazzarrini2024virtual}
E.~Gazzarrini, E.~G. Garcia, D.~Gosein, and X.~Espinal, ``The virtual research environment: A multi-science analysis platform,'' in \emph{EPJ Web of Conferences}, vol. 295.\hskip 1em plus 0.5em minus 0.4em\relax EDP Sciences, 2024, p. 08023.

\bibitem{zhao2022notebook}
Z.~Zhao, S.~Koulouzis, R.~Bianchi, S.~Farshidi, Z.~Shi, R.~Xin, Y.~Wang, N.~Li, Y.~Shi, J.~Timmermans \emph{et~al.}, ``Notebook-as-a-vre (naavre): From private notebooks to a collaborative cloud virtual research environment,'' \emph{Software: Practice and Experience}, vol.~52, no.~9, pp. 1947--1966, 2022.

\bibitem{Assante2023}
M.~Assante, L.~Candela, D.~Castelli, R.~Cirillo, G.~Coro, A.~Dell'Amico, L.~Frosini, L.~Lelii, M.~Lettere, F.~Mangiacrapa \emph{et~al.}, ``Virtual research environments co-creation: The d4science experience,'' \emph{Concurrency and Computation: Practice and Experience}, vol.~35, no.~18, p. e6925, 2023.

\bibitem{wang2013cybergis}
S.~Wang, L.~Anselin, B.~Bhaduri, C.~Crosby, M.~F. Goodchild, Y.~Liu, and T.~L. Nyerges, ``Cybergis software: a synthetic review and integration roadmap,'' \emph{International Journal of Geographical Information Science}, vol.~27, no.~11, pp. 2122--2145, 2013.

\bibitem{Yin2017}
D.~Yin, Y.~Liu, A.~Padmanabhan, J.~Terstriep, J.~Rush, and S.~Wang, ``{A cybergis-jupyter framework for geospatial analytics at scale},'' \emph{ACM International Conference Proceeding Series}, vol. Part F1287, 2017.

\bibitem{zonca2018deploying}
A.~Zonca and R.~S. Sinkovits, ``Deploying jupyter notebooks at scale on xsede resources for science gateways and workshops,'' in \emph{Proceedings of the Practice and Experience on Advanced Research Computing}.\hskip 1em plus 0.5em minus 0.4em\relax ACM, 2018, pp. 1--7.

\bibitem{lawrence2015science}
K.~A. Lawrence, M.~Zentner, N.~Wilkins-Diehr, J.~A. Wernert, M.~Pierce, S.~Marru, and S.~Michael, ``Science gateways today and tomorrow: positive perspectives of nearly 5000 members of the research community,'' \emph{Concurrency and Computation: Practice and Experience}, vol.~27, no.~16, pp. 4252--4268, 2015.

\bibitem{huang2019software}
G.~Huang, C.~Luo, K.~Wu, Y.~Ma, Y.~Zhang, and X.~Liu, ``Software-defined infrastructure for decentralized data lifecycle governance: principled design and open challenges,'' in \emph{2019 IEEE 39th International Conference on Distributed Computing Systems (ICDCS)}.\hskip 1em plus 0.5em minus 0.4em\relax IEEE, 2019, pp. 1674--1683.

\bibitem{assante2019enacting}
M.~Assante, L.~Candela, D.~Castelli, R.~Cirillo, G.~Coro, L.~Frosini, L.~Lelii, F.~Mangiacrapa, P.~Pagano, G.~Panichi \emph{et~al.}, ``Enacting open science by d4science,'' \emph{Future Generation Computer Systems}, vol. 101, pp. 555--563, 2019.

\bibitem{goble2021implementing}
C.~Goble, S.~Soiland-Reyes, F.~Bacall, S.~Owen, A.~Williams, I.~Eguinoa, B.~Droesbeke, S.~Leo, L.~Pireddu, L.~Rodr{\'\i}guez-Navas \emph{et~al.}, ``Implementing fair digital objects in the eosc-life workflow collaboratory,'' \emph{Zenodo}, 2021.

\bibitem{warnat2021swarm}
S.~Warnat-Herresthal, H.~Schultze, K.~L. Shastry, S.~Manamohan, S.~Mukherjee, V.~Garg, R.~Sarveswara, K.~H{\"a}ndler, P.~Pickkers, N.~A. Aziz \emph{et~al.}, ``Swarm learning for decentralized and confidential clinical machine learning,'' \emph{Nature}, vol. 594, no. 7862, pp. 265--270, 2021.

\bibitem{blythman2022libraries}
R.~Blythman, M.~Arshath, J.~Sm{\'e}kal, H.~Shaji, S.~Vivona, and T.~Dunmore, ``Libraries, integrations and hubs for decentralized ai using ipfs,'' \emph{arXiv preprint arXiv:2210.16651}, 2022.

\bibitem{zheng2018blockchain}
Z.~Zheng, S.~Xie, H.-N. Dai, X.~Chen, and H.~Wang, ``Blockchain challenges and opportunities: A survey,'' \emph{International journal of web and grid services}, vol.~14, no.~4, pp. 352--375, 2018.

\bibitem{zheng2017overview}
Z.~Zheng, S.~Xie, H.~Dai, X.~Chen, and H.~Wang, ``An overview of blockchain technology: Architecture, consensus, and future trends,'' in \emph{2017 IEEE international congress on big data (BigData congress)}.\hskip 1em plus 0.5em minus 0.4em\relax Ieee, 2017, pp. 557--564.

\bibitem{rong2022openiac}
C.~Rong, J.~Geng, T.~J. Hacker, H.~Bryhni, and M.~G. Jaatun, ``Openiac: open infrastructure as code-the network is my computer,'' \emph{Journal of Cloud Computing}, vol.~11, no.~1, p.~12, 2022.

\bibitem{ding2022desci}
W.~Ding, J.~Hou, J.~Li, C.~Guo, J.~Qin, R.~Kozma, and F.-Y. Wang, ``Desci based on web3 and dao: A comprehensive overview and reference model,'' \emph{IEEE Transactions on Computational Social Systems}, vol.~9, no.~5, pp. 1563--1573, 2022.

\bibitem{cao2022decentralized}
L.~Cao, ``Decentralized ai: Edge intelligence and smart blockchain, metaverse, web3, and desci,'' \emph{IEEE Intelligent Systems}, vol.~37, no.~3, pp. 6--19, 2022.

\bibitem{wang2022scaling}
Y.~Wang, S.~Koulouzis, R.~Bianchi, N.~Li, Y.~Shi, J.~Timmermans, W.~D. Kissling, and Z.~Zhao, ``Scaling notebooks as re-configurable cloud workflows,'' \emph{Data Intelligence}, vol.~4, no.~2, pp. 409--425, 2022.

\bibitem{milligan2018jupyter}
M.~B. Milligan, ``Jupyter as common technology platform for interactive hpc services,'' in \emph{Proceedings of the Practice and Experience on Advanced Research Computing}.\hskip 1em plus 0.5em minus 0.4em\relax ACM, 2018, pp. 1--6.

\bibitem{de2008myexperiment}
D.~De~Roure, C.~Goble, J.~Bhagat, D.~Cruickshank, A.~Goderis, D.~Michaelides, and D.~Newman, ``myexperiment: defining the social virtual research environment,'' in \emph{2008 IEEE fourth international conference on EScience}.\hskip 1em plus 0.5em minus 0.4em\relax IEEE, 2008, pp. 182--189.

\bibitem{kontomaris2023cwl}
C.~Kontomaris, Y.~Wang, and Z.~Zhao, ``Cwl-flops: A novel method for federated learning operations at scale,'' in \emph{2023 IEEE 19th International Conference on e-Science (e-Science)}.\hskip 1em plus 0.5em minus 0.4em\relax IEEE, 2023, pp. 1--2.

\bibitem{wang2024price}
Y.~Wang, N.~Kanwal, K.~Engan, C.~Rong, P.~Grosso, and Z.~Zhao, ``Price: Privacy-preserving and cost-effective scheduling for parallelizing the large medical image processing workflow over hybrid clouds,'' \emph{arXiv preprint arXiv:2405.15398}, 2024.

\bibitem{soiland2022packaging}
S.~Soiland-Reyes, P.~Sefton, M.~Crosas, L.~J. Castro, F.~Coppens, J.~M. Fern{\'a}ndez, D.~Garijo, B.~Gr{\"u}ning, M.~La~Rosa, S.~Leo \emph{et~al.}, ``Packaging research artefacts with ro-crate,'' \emph{Data Science}, vol.~5, no.~2, pp. 97--138, 2022.

\bibitem{barker2014schema}
P.~Barker and L.~M. Campbell, ``What is schema. org,'' \emph{LRMI. Retrieved April}, vol.~21, p. 2015, 2014.

\bibitem{schultes2019fair}
E.~Schultes and P.~Wittenburg, ``Fair principles and digital objects: Accelerating convergence on a data infrastructure,'' in \emph{Data Analytics and Management in Data Intensive Domains: 20th International Conference, DAMDID/RCDL 2018, Moscow, Russia, October 9--12, 2018, Revised Selected Papers 20}.\hskip 1em plus 0.5em minus 0.4em\relax Springer, 2019, pp. 3--16.

\bibitem{rauchs2018distributed}
M.~Rauchs, A.~Glidden, B.~Gordon, G.~C. Pieters, M.~Recanatini, F.~Rostand, K.~Vagneur, and B.~Z. Zhang, ``Distributed ledger technology systems: A conceptual framework,'' \emph{Available at SSRN 3230013}, 2018.

\bibitem{lee2019using}
W.-M. Lee and W.-M. Lee, ``Using the metamask chrome extension,'' \emph{Beginning Ethereum Smart Contracts Programming: With Examples in Python, Solidity, and JavaScript}, pp. 93--126, 2019.

\bibitem{benet2014ipfs}
J.~Benet, ``Ipfs-content addressed, versioned, p2p file system,'' \emph{arXiv preprint arXiv:1407.3561}, 2014.

\end{thebibliography}
